\documentclass{appolb}
\usepackage{epsfig}
\begin{document}
%\date{\today}

\newcommand{\drbar}{{\overline{DR}}}
\newcommand{\msbar}{{\overline{MS}}}
\newcommand{\GeV}{{\rm GeV}}
\newcommand{\stau}{\tilde{\tau}}
\newcommand{\snt}{{\tilde{\nu}_\tau}}
\newcommand{\ur}{\tilde{u}_R}
\newcommand{\ul}{\tilde{u}_L}
\newcommand{\dr}{\tilde{d}_R}
\newcommand{\dl}{\tilde{d}_L}
\newcommand{\st}{\tilde{t}}
\newcommand{\sbot}{\tilde{b}}
\newcommand{\sg}{\tilde{g}}
\newcommand{\nt}{\tilde{\chi}^0}
\newcommand{\cp}{\tilde{\chi}^+}
\newcommand{\cm}{\tilde{\chi}^-}
\newcommand{\cx}{\tilde{\chi}}
\newcommand{\ser}{\tilde{e}_R}
\newcommand{\sel}{\tilde{e}_L}
\newcommand{\sne}{\tilde{\nu}_e}
\def\babar{\mbox{\sl B\hspace{-0.4em} {\scriptsize\sl A}\hspace{-0.37em} \sl
    B\hspace{-0.4em} {\scriptsize\sl A\hspace{-0.02em}R}}}
\def\simlt{\stackrel{<}{{}_\sim}}
\def\simgt{\stackrel{>}{{}_\sim}}

\pagestyle{plain}
%% uncomment the following line to get equations numbered by (sec.num)
%\eqsec
\newcount\eLiNe\eLiNe=\inputlineno\advance\eLiNe by -1
\title{Looking for CP violation in $B\to \tau^+\tau^-$ decays%
\thanks{Talk given at the International Conference ``Matter to the Deepest'',
  Ustro\'n, Poland, 8-14 September 2005}%
}
\author{J. KALINOWSKI$^a$\thanks{Speaker. E-mail address:
Jan.Kalinowski@fuw.edu.pl.}, P.H. CHANKOWSKI$^a$, Z. WAS$^b$ 
and~M.~WOREK$^{b,c}$ % 
\address{$^a$Institute of Theoretical Physics, Warsaw University, 
00-681~Warsaw, Poland\\ $^b$Institute of Nuclear Physics PAS, 
31-342 Cracow, Poland\\
$^c$Institute of Nuclear Physics, NCRS ``Demokritos'', 15310 Athens, Greece }}
\maketitle

\begin{abstract}
In 
supersymmetry with large $\tan\beta$ the decays $B^0(\bar{B}^0)\rightarrow
l^+l^-$ are dominated by 
the scalar and pseudoscalar Higgs penguin diagrams leading to strong
enhancement of leptonic decay rates with potentially   
large CP asymmetries in the $\tau^+\tau^-$ decay modes  
measurable in BELLE or \babar\ experiments. 
The {\tt TAUOLA}
$\tau$-lepton decay library supplemented by its universal interface can 
efficiently be used to search for $B^0(\bar{B}^0)\rightarrow\tau^+\tau^-$ 
decays, and  to investigate how the CP asymmetry is reflected in realistic 
experimental observables.

\end{abstract}

\section{Introduction}
Understanding the origin of CP violation is one of the most important tasks of
particle physics. Until now, CP violation has been firmly established in $K$-
and $B$-physics in a series of high statistics experiments.  One way of
testing the conventional CKM description of CP violation is by more precise
measurements and joint analysis of CP asymmetries measured in different
channels with a hope of finding some inconsistency signaling a contribution
from physics beyond the Standard Model (SM) to CP violation.  Alternatively,
observation of a non-zero effect in channels in which no (or negligibly small)
effects violating CP are predicted by the SM would be an unambiguous signal of
new physics contribution to CP violation.

In this talk presented are results of recent analyses of flavour changing
decays of the neutral $B$ mesons into lepton pairs, $B^0_{d,s}\rightarrow
l^+l^-$ \cite{Chankowski:2004tb}, for which the SM predicts no CP violating
effects. This decay mode is very sensitive to new physics which affects the
$b$-quark Yukawa couplings \cite{KASKHAPOTO,BAKO,CHSL}.  Approximate and full
one-loop calculations in the supersymmetric extension of the SM with large
$\tan\beta$ (the ratio of the vacuum expectation values of the two Higgs
doublets) \cite{BAKO,CHSL,ISRE} showed that truly spectacular enhancement of
the rates of the decays $B^0_{d,s}\rightarrow l^+l^-$ can be expected.
Moreover, new physics can also lead to observable CP violation in these
decays~\cite{Chankowski:2004tb}.  CP violation could manifest itself through
non-equal leptonic decay rates of the $B^0(t)$ and $\bar B^0(t)$ states
(tagged $t=0$ as $B^0$ and $\bar B^0$, respectively). If polarization of final
state leptons can be determined, additional information on CP violation could
be provided by non-equal $\Gamma(B^0(t)\rightarrow l^+_Ll^-_L)$ and
$\Gamma(\bar B^0(t)\rightarrow l^+_Rl^-_R)$ [or $\Gamma(B^0(t)\rightarrow
l^+_Rl^-_R)$ and $\Gamma(\bar B^0(t)\rightarrow l^+_Ll^-_L)$] decay rates
\cite{Huang:2000tz,Dedes:2002er}.

Leptonic decays are theoretically  clean as the only
non-perturbative quantities they depend on, are the $B^0$
meson decay constants $F_{B_{d,s}}$, which cancel
out in suitably defined CP asymmetries. However, none of 
these decays have been seen so far: the
best upper limits on the $B^0_{d,s} \rightarrow\mu^+\mu^-$ and $\tau^+\tau^-$ 
branching
fractions at present are listed in Table~\ref{tab:BR}.

\begin{table}[t] 
\begin{center}$
\begin{array}{|c|r|rl|}
\hline
\mbox{Mode} & \mbox{SM expectation} & \mbox{Exp.\ limit}& \mbox{~~~~Ref.}\\ 
\hline 
B_d\to \mu^+\mu^-   & 1\times 10^{-10} & 8.3\times 10^{-8}& 
\mbox{~~\cite{BaBarICHEP04}} \\
&& 3.8\times 10^{-8} & \mbox{~~\cite{CDFSUSY05}}\\
{B_d\to \tau^+\tau^-} &{2.8\times 10^{-8}} & {2.7\times
  10^{-3}}& 
{ \mbox{~~\cite{BaBarEPS05}}}\\
B_s\to \mu^+\mu^- & 3.7 \times 10^{-9} & 1.5\times 10^{-7}& 
\mbox{~~\cite{CDFSUSY05}}\\ 
{B_s\to \tau^+\tau^-} & { 1\times 10^{-6}} & { \sim 5\times
  10^{-2}}  & {  \mbox{~~\cite{LEP96}}} \\
\hline
\end{array}$
\
\end{center}
\caption{Upper limits for the branching fractions of leptonic $B$-meson
  decays.} \label{tab:BR} 
\end{table}

With the rates as
predicted by the SM, the detection of the $\mu^+\mu^-$ decay channel 
will become possible only at the LHC; in the
hadronic collider the $\tau^+\tau^-$ channel is extremely challenging. 
New physics (like supersymmetry) can increase significantly their rates to 
a level that they can be observed at \babar, BELLE or Tevatron in near 
future. We find, however, that the ratio of time 
integrated leptonic decay rates of $B^0(t)\to\mu^+\mu^-$ and
$\bar{B}^0(t)\to\mu^+\mu^-$  
is  unlikely to deviate appreciably from unity. Polarization 
measurement also seems very difficult in the case of the $\mu^+\mu^-$
channel. 
In the $\tau^+\tau^-$ channel the situation can be quite 
different: large CP violating effects can be expected, and $\tau$ 
polarization measurement is possible.  Results of 
first  experimental analyses of this
channel have recently been made public~\cite{BaBarEPS05}. Increasing the 
experimental efficiency of identifying $\tau$ 
leptons and suppressing the background might result in a much stronger limit
or, hopefully, in observing the signal in $B$ meson factories. We show that 
the existing {\tt TAUOLA} package 
\cite{tauola} and its 
{\tt universal interface}  \cite{interface} 
may prove useful to search for these decays {\it and} for the CP violation 
since  in  
realistic scenarios of large $\tan\beta$ MSSM the decay rates can
significantly be increased 
to a level measurable at the running \babar\ and BELLE experiments and  the  
CP asymmetry in the $B^0_d(\bar{B}^0_d)\to\tau^+\tau^-$ channel can be 
quite large  and potentially measurable.
We also identify,  in addition to the ratio of time integrated leptonic
$B^0_d(t)$ and  
$\bar B^0_d(t)$, two    realistic CP-sensitive  
experimental observables which reflect $\tau$-lepton polarisations:
the $\pi^\pm$ energies from $\tau\to\pi\nu$ decays and 
the acoplanarity angle between
the decay planes 
of the $\rho$ mesons which originate from $\tau\to\rho\nu$. The
former is sensitive to the longitudinal, while the latter to the
transverse polarizations of $\tau$'s  coming from $B^0_d$ and
$\bar B^0_d$ decays. 

\section{Preliminaries}\label{sec:formulas}
%%%%%%%%%%%%%%%%%%%%%%%%%%%%%%%%%%%%%%%%%%%%%%%%%%%%%%%%%%%%%%%%%%%%%%%
In leptonic $B$-meson decays  if, for
example,  
\begin{eqnarray}
\Gamma(B^0\rightarrow l^+_Ll^-_L)\neq \Gamma(\bar{B}^0\rightarrow l^+_Rl^-_R)~,
\label{eqn:cpt_viol1} \nonumber\\
\Gamma(B^0\rightarrow l^+_Rl^-_R)\neq \Gamma(\bar{B}^0\rightarrow l^+_Ll^-_L)~
\label{eqn:cpt_viol2}
\end{eqnarray}
CP is violated 
because the initial and final states on both sides transform into each other 
under CP \cite{Huang:2000tz}.
The amplitudes of $B^0$ decays into  two helicity eigenstates read
\begin{eqnarray}
{\cal A}_L\equiv\langle l^+_Ll^-_L|B^0\rangle
=M_B\left(a + b~\beta\right)~,\nonumber\\
{\cal A}_R\equiv\langle l^+_Rl^-_R|B^0\rangle
=M_B\left(a - b~\beta\right)~,\label{eqn:ALRdefs}
\end{eqnarray}
Similar formulae with $a$ and $b$ 
replaced by $\bar a$ and $\bar b$, respectively, give the amplitudes
$\bar{\cal A}_L$ and $\bar{\cal A}_R$ for the corresponding $\bar{B}^0$ decays.
Here $\beta=(1-4m_l^2/M^2_B)^{1/2}$ and $a$, $b$, $\bar a$ and $\bar b$  
are the coefficients in the effective Lagrangian 
describing $B^0(\bar{B}^0)\rightarrow l^+l^-$ decays 
\begin{eqnarray}
{\cal L}_{\rm eff} = B^0_{s,d}\bar\psi_l(b_{s,d}+a_{s,d}\gamma^5)\psi_l
         + \bar{B}^0_{s,d}\bar\psi_l(\bar b_{s,d}+\bar a_{s,d}\gamma^5)\psi_l~.
\label{eqn:leff}
\end{eqnarray}
(the subscripts $d$ and $s$ referring to 
non-strange and strange $B^0$ mesons, unless explicitly written,  
will be omitted).
Hermiticity (CPT invariance) implies
$\bar b = b^\ast$ and $\bar a = -a^\ast$.

Since in leptonic decays no strong phases are involved, 
inequality (\ref{eqn:cpt_viol2})  
can occur only through the mixing of the $B^0$ and $\bar{B}^0$ mesons. In the 
standard formalism \cite{BLS} the state which at $t=0$ is a pure $B^0$
($\bar{B}^0$)  
evolves in time as
\begin{eqnarray} 
|B^0_{\rm phys}(t)\rangle=g_+(t)|B^0\rangle+{q\over p}g_-(t)|\bar{B}^0\rangle~
\end{eqnarray}
(for $\bar B^0_{\rm phys}(t)$ replace $B^0\leftrightarrow\bar B^0$ and 
$p\leftrightarrow q$). The coefficients $g_{\pm}(t)$,  neglecting the
difference of the decay widths of 
the two $B^0$ mass  eigenstates and denoting 
$\Delta M\equiv M_{B_H^0}-M_{B_L^0}\ll
M_B\equiv(M_{B_H^0}+M_{B_L^0})/2$, read  
\begin{eqnarray} 
g_+(t) = e^{-iM_Bt-{\Gamma\over2}t} ~\cos{\Delta M\over2} t~, 
\phantom{aa}%\nonumber\\
g_-(t) = e^{-iM_Bt-{\Gamma\over2}t} ~i ~\sin{\Delta M\over2} t~, 
\end{eqnarray}
and  the ratio $p/q$ is calculated from the effective Hamiltonian 
\begin{eqnarray} 
{p/ q}=\left({{H_{12}^\ast/ H_{12}}}\right)^{1/2}
%\approx{M_{12}^\ast\over |M_{12}|}\left(1-{1\over2}{\rm Im}
%{\Gamma_{12}\over M_{12}}\right)~,
\end{eqnarray}
with $H_{12}\equiv M_{12}+{i\over2}\Gamma_{12}=
\langle B^0|{\cal H}_{\rm eff}|\bar{B}^0\rangle$, {\it etc.} \cite{BLS}.

Since $|{\cal A}_L|=|\bar {\cal A}_R|$, $|{\cal A}_R|=|\bar {\cal A}_L|$
and 
$ 
{\bar{\cal A}_L/{\cal A}_L}=
\left({{\cal A}_R/\bar{\cal A}_R}\right)^\ast,
$
CP is violated if  either
\begin{eqnarray} 
\left|{q/ p}\right|\neq1\phantom{aaaaaa}{\rm or}
\phantom{aaaaaa}{\rm Im}(\lambda_L)\neq {\rm Im}(\lambda_R^{-1})~,\nonumber
\end{eqnarray}
where 
$
\lambda_L\equiv q\, \bar{\cal A}_L/p\, {\cal A}_L$ 
and 
$\lambda_R\equiv q\, \bar{\cal A}_R/p\, {\cal A}_R~.
$

The simplest quantitative measures of CP violation
are provided by the asymmetries constructed out of time 
integrated polarized 
decay rates
\begin{eqnarray} 
&&A_{\rm CP}^1(t_1,t_2)\equiv
{\int_{t_1}^{t_2} dt~[\Gamma(B^0_{\rm phys}(t)\rightarrow l^+_Ll^-_L)
- \Gamma(\bar{B}^0_{\rm phys}(t)\rightarrow l^+_Rl^-_R)]\over
\int_{t_1}^{t_2} dt~[\Gamma(B^0_{\rm phys}(t)\rightarrow l^+_Ll^-_L)
+\Gamma(\bar{B}^0_{\rm phys}(t)\rightarrow l^+_Rl^-_R)]}~,
\label{eqn:as1cp}
\\[1mm]
&&A_{\rm CP}^2(t_1,t_2)\equiv
{\int_{t_1}^{t_2} dt~[\Gamma(B^0_{\rm phys}(t)\rightarrow l^+_Rl^-_R)
-\Gamma(\bar{B}^0_{\rm phys}(t)\rightarrow l^+_Ll^-_L)]\over
\int_{t_1}^{t_2} dt~[\Gamma(B^0_{\rm phys}(t)\rightarrow l^+_Rl^-_R)
+\Gamma(\bar{B}^0_{\rm phys}(t)\rightarrow l^+_Ll^-_L)]}~.
\label{eqn:as2cp}
\end{eqnarray}
and the ratio of integrated unpolarized decay rates
\begin{eqnarray} 
R_l(t_1,t_2)\equiv
{\int_{t_1}^{t_2} dt~\Gamma(B^0_{\rm phys}(t)\rightarrow l^+l^-)\over
\int_{t_1}^{t_2} dt~\Gamma(\bar B^0_{\rm phys}(t)\rightarrow l^+l^-)}~.
\label{eqn:R_l}
\end{eqnarray}
If the statistics of  tagged events is low, or 
experimental determination 
of the decay time $t$ is difficult, the fully integrated asymmetries
$A_{\rm CP}^1\equiv A_{\rm CP}^1(0,\infty)$, 
$A_{\rm CP}^2\equiv A_{\rm CP}^2(0,\infty)$ and the ratio 
$R_l\equiv R_l(0,\infty)$ can be exploited. 
If $|q/p|=1$ (as in the SM and many SUSY extensions), 
the expressions for the 
asymmetries $A_{\rm CP}^1$, $A_{\rm CP}^2$ and the ratio 
$R_l$ 
simplify to
\begin{eqnarray} 
&&A_{\rm CP}^1=
{-2~x~{\rm Im}~\lambda_L\over2+x^2+x^2~|\lambda_L|^2}~,\phantom{aaaaa}
A_{\rm CP}^2=
{-2~x~{\rm Im}\lambda_R\over2+x^2+x^2~|\lambda_R|^2}~, \label{eq:Asimpl}
\\[1mm]
%\end{eqnarray}
%\begin{eqnarray} 
&& R_l={\left(|{\cal A}_L|^2+|{\cal A}_R|^2\right)(1+x^2)
-x\left\{|{\cal A}_L|^2{\rm Im}(\lambda_L)
+|{\cal A}_R|^2{\rm Im}(\lambda_R)\right\}\over
\left(|{\cal A}_L|^2+|{\cal A}_R|^2\right)(1+x^2)
+x \left\{|{\cal A}_L|^2{\rm Im}(\lambda_L)
+|{\cal A}_R|^2{\rm Im}(\lambda_R)\right\}}.\mbox{\phantom{mm}}
\label{eqn:Rlsimpl}
\end{eqnarray}
where
$ x\equiv \Delta M/\Gamma$. 
The asymmetries $A_{\rm CP}^1$, $A_{\rm CP}^2$, as functions of
$\lambda_{L,R}$, are therefore bounded from above by \cite{Huang:2000tz}
\begin{eqnarray} 
\left|A_{\rm CP}^{1,2}\right|\leq (2+x^2)^{-1/2}~.
\end{eqnarray}
which has important consequences.
Since $x_s>20.6$ for the $B^0_s$-$\bar{B}^0_s$ system, the CP asymmetries
in the leptonic $B^0_s(\bar{B}^0_s)$ decays can reach at best~$\sim$~4.5\%
irrespectively of the amount of CP violation.
In contrast, for the $B^0_d$-$\bar{B}^0_d$ system, for which 
$x_d=0.771\pm0.012$ they can be as large as~$\sim$~60\% and hopefully
measurable  at \babar\ and BELLE in a 
relatively clean environment. Therefore below  
we will consider only the CP asymmetries
in the $B^0_d(\bar{B}^0_d)\rightarrow l^+l^-$ decays.

The time dependent $B^0$-meson mixing can easily be 
dealt with by introducing time dependent effective factors $a_{\rm eff}$, 
$b_{\rm eff}$, $\bar a_{\rm eff}$ and $\bar b_{\rm eff}$. The  factor $a_{\rm
  eff}$ is 
defined as
\begin{eqnarray} 
a_{\rm eff}(t) = a ~g_+(t) + \bar a ~{q\over p}~g_-(t)
\end{eqnarray}
and the other factors are defined in a similar manner~\cite{Chankowski:2004tb}. 
Then the instantaneous $B^0$ widths into left- or 
right-handed $l$'s read
\begin{eqnarray} 
\Gamma(B^0_{\rm phys}(t)\rightarrow l^+_L l^-_L\, 
[l^+_R l^-_R])={M_B\over16\pi}\beta
\left|a_{\rm eff}(t) + [-] \beta ~b_{\rm eff}(t)\right|^2
\end{eqnarray}
and those for $\bar{B}^0$ are given by similar formulae with 
$a_{\rm eff}(t),~b_{\rm eff}(t)$ replaced by 
$\bar a_{\rm eff}(t),~\bar b_{\rm eff}(t)$. CP is violated because in general  
$\bar a_{\rm eff}(t)\neq-a_{\rm eff}^\ast(t)$, and
$\bar b_{\rm eff}(t)\neq b_{\rm eff}^\ast(t)$.

In  the case of the $\tau^+\tau^-$ decay mode the $\tau$ polarisations    
are best identified by  measuring  the $\pi^\pm$ energy 
spectra from $\tau\rightarrow\pi\nu$ decays \cite{taupol}. 
The spin density 
matrix formalism of ref.~\cite{Desch:2003rw} 
allows to construct also observables 
sensitive to transverse polarization of the final state $\tau$'s.  
The spin weight for the complete event ($B\to\tau\tau\to$ decay
products)  can be written in  the form 
%(for details we refer the reader to \cite{Jadach:1990mz})
\begin{eqnarray}
WT={1\over4}(1+\sum_{i,j=x,y,z}R_{ij}~h^i_1~h^j_2
+\sum_{i=x,y,z}R_{i0}~h^i_1+\sum_{j=x,y,z}R_{0j}~h^j_2
)~.\label{eqn:spwgh}
\end{eqnarray}
where the polarimetric vectors $\vec{h}_1$ 
and $\vec{h}_2$ are determined   
solely by the dynamic of the $\tau$ decay processes, and  
\begin{eqnarray}
\label{eq:matrixR}
&& R_{00}=+1, \qquad \qquad R_{x0}=R_{y0}=R_{0x}=R_{0y}=0, \nonumber \\
&& R_{zz}=-1, \qquad \qquad R_{xz}=R_{yz}=R_{zx}=R_{zy}=0,  \nonumber \\
&&R_{0z}=-R_{z0} = {2{\rm Re}(a_{\rm eff}b^\ast_{\rm eff})~\beta
\over|b_{\rm eff}|^2\beta^2+|a_{\rm eff}|^2},\\
&&R_{xy}=-R_{yx}=-{2{\rm Im}(a_{\rm eff}b^\ast_{\rm eff})~\beta
\over|b_{\rm eff}|^2\beta^2+|a_{\rm eff}|^2}, \phantom{aaaa}
R_{xx}=R_{yy}={|b_{\rm eff}|^2\beta^2-|a_{\rm eff}|^2
\over|b_{\rm eff}|^2\beta^2+|a_{\rm eff}|^2},\nonumber
\end{eqnarray}
Eq.~(\ref{eqn:spwgh}) with $R_{\mu\nu}$ replaced by 
$\bar R_{\mu\nu}$
computed as above but with $\bar a_{\rm eff}(t)$ and $\bar b_{\rm eff}(t)$
replacing $a_{\rm eff}(t)$ and $b_{\rm eff}(t)$, respectively, gives the spin 
weight for the events from $\bar{B}^0$ decays.  
CP violating effects in the $B^0\rightarrow\tau^+\tau^-$ and
$\bar B^0\rightarrow\tau^+\tau^-$ decays
are absent if $R_{0z}=\bar R_{z0}$,  
$R_{xy}=\bar R_{yx}$ and/or  
$R_{xx}=\bar R_{xx}$.

To simulate how the CP asymmetry in
$B_d^0(\bar{B}_d^0)\rightarrow\tau^+\tau^-$ decays are reflected in realistic
observables the {\tt TAUOLA} $\tau$-lepton decay library has been used (the
method and numerical algorithm is given in \cite{tauola}).  The input to the
{\tt TAUOLA Universal Interface} \cite{interface} is the spin density matrix
of the $\tau^+\tau^-$ system resulting from the decay of a neutral particle.

For simulations of the time integrated measurements, the time 
averaged matrix $\langle R_{\mu\nu}\rangle$ has to be used
\begin{eqnarray}
\langle R_{\mu\nu}\rangle\equiv
{\int dt ~\Gamma(B^0_{\rm phys}(t)\rightarrow \tau^+\tau^-) ~
R_{\mu\nu}(t)
\over\int dt ~\Gamma(B^0_{\rm phys}(t)\rightarrow \tau^+\tau^-)}
\nonumber
\end{eqnarray}
and $\langle\bar R_{\mu\nu}\rangle$ given by a similar formula.
The asymmetry (\ref{eqn:as1cp}) is then given by
\begin{eqnarray}
A_{\rm CP}^1={(1-\langle R_{z0}\rangle)\Gamma_{\rm int}
-(1+\langle\bar R_{z0}\rangle)\bar\Gamma_{\rm int}\over
(1-\langle R_{z0}\rangle)\Gamma_{\rm int} 
+(1+\langle\bar R_{z0}\rangle)\bar\Gamma_{\rm int}}~,
\label{eqn:asymbyR}
\end{eqnarray}
where
\begin{eqnarray}
\Gamma_{\rm int}=
\int dt ~\Gamma(B^0_{\rm phys}(t)\rightarrow
\tau^+\tau^-), 
\phantom{aaa}
\bar\Gamma_{\rm int}=
\int dt ~\Gamma(\bar B^0_{\rm phys}(t)\rightarrow \tau^+\tau^-).
\nonumber
\end{eqnarray}
$A_{\rm CP}^2(t_1,t_2)$ defined in (\ref{eqn:as2cp}) 
is given by (\ref{eqn:asymbyR}) reversing the signs in the brackets.

In the limit $a=\pm b\beta$ the two integrated rates $\Gamma_{\rm int}$ and
$\bar\Gamma_{\rm int}$ are equal resulting in $\langle R_{z0}\rangle=
-\langle\bar R_{z0}\rangle$. It implies that even for $\delta_{\rm CP}\neq0$
there can be no CP violating effects in the observables sensitive to the
longitudinal polarization of $\tau$'s, like $A_{\rm CP}^{1,2}$, nor in
$R_\tau$.  However, in this limit the elements $xx$ and $xy$ of these matrices
{\it need not} satisfy $\langle R_{xy}\rangle= -\langle\bar R_{xy}\rangle$ and
$\langle R_{xx}\rangle= \langle\bar R_{xx}\rangle$.  Hence, the observables
sensitive to transverse $\tau$ polarization can reveal CP violation even if
the observables sensitive to longitudinal $\tau$ polarization cannot.

\section{Supersymmetry scenarios}\label{sec:susy}
%%%%%%%%%%%%%%%%%%%%%%%%%%%%%%%%%%%%%%%%%%%%%%%%%%%%%%%%%%%%%%%%%%%%%%%%

In models of new physics ${\rm Im}(\lambda_L)\neq{\rm Im}(\lambda_R^{-1})$
and/or $|q/p|\neq1$ can be expected. However, if the decay rates are not at
the same time strongly enhanced, the detection of the $\mu^+\mu^-$ decay mode
will become possible only at the LHC, while for the
$B^0_d\rightarrow\tau^+\tau^-$ mode at BELLE and \babar\ the number of
reconstructed events might be too small to detect any CP violation.

Much more promising situation can occur in the supersymmetric scenario with a
large ratio of the vacuum expectation values of the two Higgs doublets,
$v_u/v_d\equiv\tan\beta\sim40\div50$.  Contributions from the Higgs penguin
diagrams with $s$-channel $H^0$ and $A^0$ Higgs boson exchanges
\cite{BAKO,CHSL,Dedes:2002er} can become dominant easily saturating the
experimental limits in Table~\ref{tab:BR} if the Higgs particles $H^0$ and
$A^0$ in the range of order $\simlt500$ GeV even for other supersymmetric
particle masses quite large, say in the TeV range \cite{CHSL,BUCHROSL}.

In Ref.~\cite{Chankowski:2004tb} two different supersymmetric
scenarios have been considered: 
minimal (MFV) and non-minimal (NMFV) flavour violating, both
with large ratio of VEVs.
%*******************************************************************  
\begin{figure}[!ht]
\begin{center}  
\epsfig{file=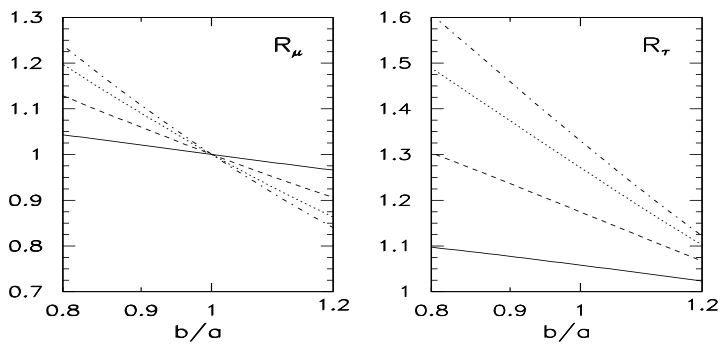,width=100mm,height=45mm}
\end{center} 
\caption  
{\it The ratios $R_\mu$ and $R_\tau$  as functions of 
$b/a$ for the phase $\delta_{\rm CP}=-{1\over2}{\rm arg}(\lambda_L)= 
0.1$ (solid line), $0.3$ (dashed), $0.5$ (dotted) and $0.75$ (dash-dotted). 
$R_l(-\delta_{\rm CP})=R^{-1}_l(\delta_{\rm CP})$.} 
\label{fig:Rl}  
\end{figure}  
%*******************************************************************
In both scenarios in which the $B^0_{d,s}\rightarrow l^+l^-$ amplitudes have
been dominated by the exchange of $H^0$ and $A^0$ Higgs bosons, it was found
that
\begin{eqnarray} 
a\approx b\phantom{aaa} {\rm or}\phantom{aaa} a\approx-b~, \nonumber
\end{eqnarray}
(up to $\simlt15\%$).
{}For $a=b$ the factors $\lambda_L$ and $\lambda_R$ simplify to:
\begin{eqnarray} 
\lambda_L=-{q\over p}~{a^\ast\over a}{1-\beta\over1+\beta}~,
\phantom{aaaaaa}
\lambda_R=-{q\over p}~{a^\ast\over a}{1+\beta\over1-\beta}~
\label{eqn:lambdas}
\end{eqnarray}
and all CP-sensitive quantities could be expressed in terms of one effective 
phase which can be taken as
\begin{eqnarray}
\delta_{\rm CP}=-{1\over2}{\rm arg}(\lambda_L)~ \label{eq:CPphase}
\end{eqnarray}
The effective CP phase is a function of the 
CKM phase and complex soft SUSY breaking Lagrangian 
parameters.  

The immediate consequence of $a=b$ with $|q/p|\sim 1$ is that for the
$\mu^+\mu^-$ final state, for which $\beta=(1-4m_\mu^2/M^2_B)^{1/2}$ is almost
1, the parameters $|\lambda_L|$ and $|\lambda_R|$ assume values
$\sim4\times10^{-4}$ and $\sim2.5\times10^3$, respectively.  As a result, the
expected asymmetries are very small, at most $|A_{\rm
  CP}^1|\simlt2\times10^{-4}$, $|A_{\rm CP}^2|\simlt10^{-3}$.
%, and  the deviation of $R_\mu$ from unity is also strongly suppressed. 
%The same is true if $a$ and $b$ are somewhat split. 
On the other hand, for the $\tau^+\tau^-$ final states 
$\beta=(1-4m_\tau^2/M^2_B)^{1/2}$ differs 
substantially from $1$ giving  $|\lambda_L|\sim0.15$, $|\lambda_R|\sim6.7$
and the maximal possible values of the asymmetries are 
\begin{eqnarray} 
\left|(A_{\rm CP}^1)^{\rm max}\right|=9\% \phantom{aaa} {\rm and}\phantom{aaa} 
\left|(A_{\rm CP}^2)^{\rm max}\right|=35\%~.
\end{eqnarray}
This is reflected in Fig.~\ref{fig:Rl} in which possible CP violating effects
in the ratio  
(\ref{eqn:Rlsimpl}) 
for  $\mu^+\mu^-$ and $\tau^+\tau^-$ decay modes are shown as functions of 
$b/a$ for four different values of the phase $\delta_{\rm CP}$ (keeping 
arg$(a)=$arg$(b)$ and $|p/q|=1$). The plots demonstrate that the ratios $R_l$ 
approach unity for $a\approx\pm b\beta$, the feature which can be read  
also from the formula (\ref{eqn:Rlsimpl}) 
if one takes into account that  for $a\to \pm b\beta$
$\lambda_R\sim1/(a\mp b\beta)$ whereas $|{\cal A}_R|^2\sim|a\mp b\beta|^2$.
Therefore, for $a\approx\pm b$ the deviation from unity of $R_\mu$ is tiny 
while for $R_\tau$ it can be quite substantial.

The coefficients $a$ and $b$ ($\bar a$ and $\bar b$) are constrained by the 
experimental limit in Table~\ref{tab:BR}, which in the case $a\approx 
\pm b$ gives
\begin{eqnarray}
|a|\approx|b|\simlt4.9\times10^{-9}~.\label{eqn:dirlim1}
\end{eqnarray}
In our simulations in the next  section 
we conservatively set $|a|=|b|\simlt10^{-9}$ and treat 
both scenarios simultaneously, as all what matters are the values of $|a|=|b|$ 
and the single CP violating phase $\delta_{\rm CP}$ which can be of order 1.
As supersymmetric box and $Z^0$ penguin contributions as well as 
a finite difference of $A^0$ and $H^0$ masses
spoil the exact equality 
$a=\pm b$, we investigate the effect of  $a$ and $b$ different from
each other by some $15\div20$\%.

\section{Numerical results}\label{sec:MC}

Two  observables, known to 
provide valuable and complementary 
information on the spin state of decaying $\tau$
lepton pairs, have been simulated: \\
(a)~ $\pi^\pm$ {\it energy spectra}  in the decay channels  
$\tau^+\to\pi^+\bar\nu_\tau$ (or $\tau^-\to\pi^-\nu_\tau$)   
reflect the longitudinal polarization of the individual $\tau^\pm$ leptons.  
Therefore they are sensitive to 
$R_{z0}$ and $\bar R_{0z}$, as can be inferred from the expression 
(\ref{eqn:spwgh}), {\it i.e.} to 
${\rm Re}(a_{\rm eff}b^\ast_{\rm eff})$ as follows from (\ref{eq:matrixR}). 
The CP violation is signaled if   
the energy spectrum of $\pi^-(\pi^+)$ originating from $B^0(\bar{B}^0)$ is
different from  
the energy spectrum of $\pi^+(\pi^-)$ originating from $B^0(\bar{B}^0)$.
In the following plots, the energy spectra are measured  
in the rest frame of the $B^0(\bar{B}^0)$ meson assuming that the 
reconstruction of the event kinematics in the BELLE and \babar\ experiments is 
sufficiently good for that purpose. \\
(b)~ {\it acoplanarity angle} $\varphi^\ast$~   
between two planes spanned by the momenta of decay products of 
$\rho^\pm\to\pi^\pm\pi^0$ coming from decays of both $\tau$ leptons into 
$\rho\nu_\tau$ \cite{Desch:2003rw}. 
It is sensitive to  
correlations between transverse components of $\tau$-lepton spins 
({\it i.e.} to the elements $R_{xx}$ and $R_{xy}$ which in turn 
probe ${\rm Im}(a_{\rm eff}b^\ast_{\rm eff})$, as can be seen from 
(\ref{eq:matrixR})).  The acoplanarity angle is defined as
\begin{eqnarray}
\varphi^\ast=\left\{ \begin{tabular}{cl}
                  $\xi$ &  ~~~~for~~ sgn$({\bf p}_{\pi^-} \cdot
                  {\bf n}_+)<0$ \\
                  $2\pi-\xi$ & ~~~~for~~ sgn$({\bf p}_{\pi^-} \cdot
                  {\bf n}_+)>0$
                  \end{tabular} \right.
\end{eqnarray}
where $\cos\xi ={{\bf n}_+ \cdot{\bf n}_-\over|{\bf n}_+||{\bf n}_-|}$ 
and two 
vectors ${\bf n}_\pm={\bf p}_{\pi^\pm}\times{\bf p}_{\pi^0}$ 
are normal to the planes determined by the momenta
of pions which originate from $\rho^\pm$ decays. 
Note that   
the full range of the variable $0 <\varphi^\ast<2\pi$ of physical 
interest, and  
in addition events have to be sorted depending whether    
$y_1 y_2>0$ or $y_1 y_2<0$, where   
\begin{equation}  
y_1={E_{\pi^+}-E_{\pi^0}\over E_{\pi^+}+E_{\pi^0}}~,\phantom{aaaaaaa}
y_2={E_{\pi^-}-E_{\pi^0}\over E_{\pi^-}+E_{\pi^0}}~,
\label{E-zone}   
\end{equation}
since otherwise the spin correlations are washed out, as explained in 
Ref.~\cite{Bower:2002zx}. The 
acoplanarity distribution is evaluated in the rest frame of the 
$\rho^+\rho^-$ pair, but with 
the energies of $\pi^\pm$ and $\pi^0$'s in (\ref{E-zone}) taken 
in the rest frame of the $B^0$($\bar B^0$). 
The difference between the distributions of the acoplanarity angle 
$\varphi^\ast$ measured in $B^0$ decays and  the angle $2\pi-\varphi^\ast$ 
measured in $\bar B^0$ decays for the same signs of $y_1y_2$ signals 
the CP violation.

%*******************************************************************  
\begin{figure}[!ht]
\begin{center}  
\epsfig{file=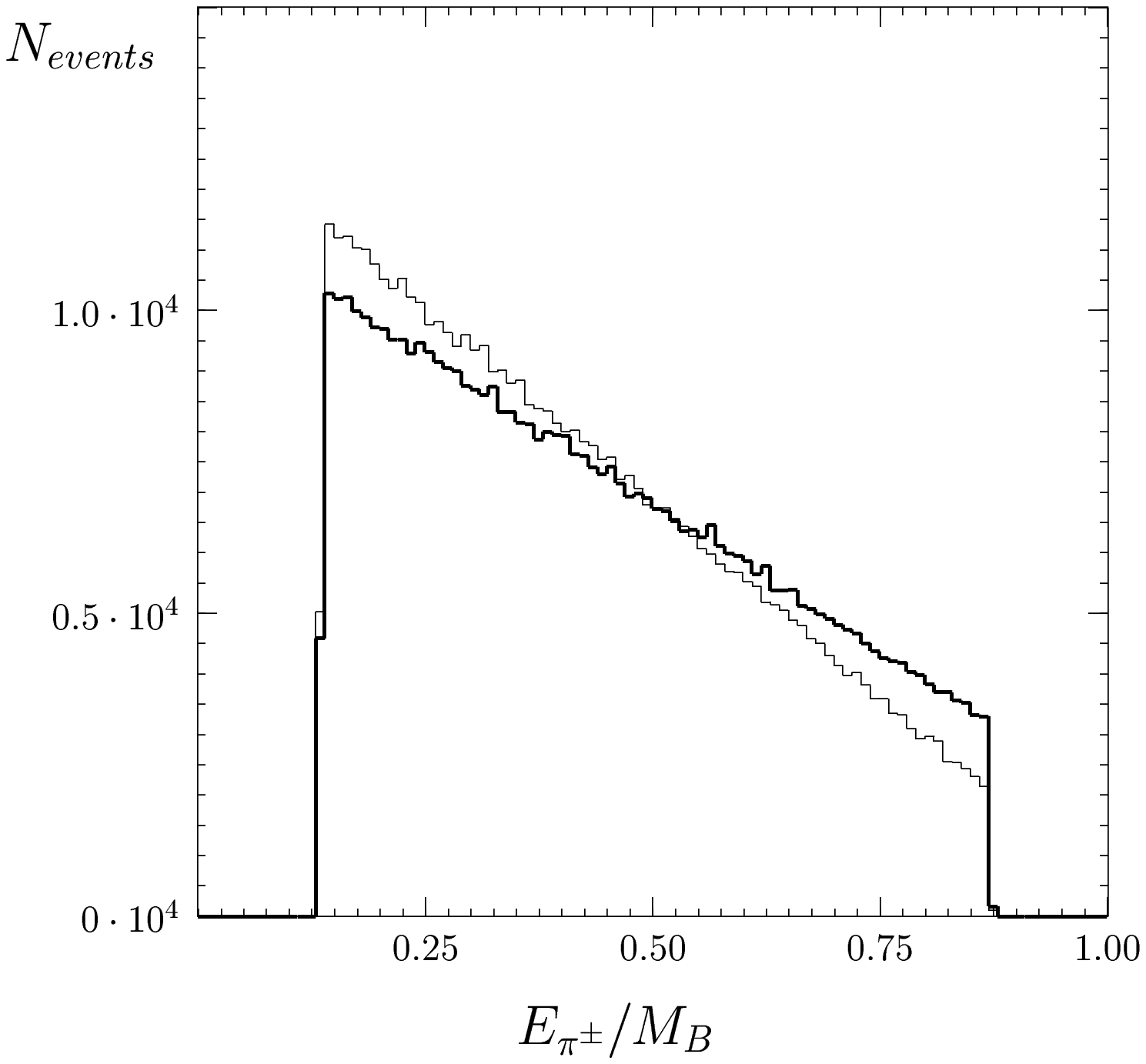,width=55mm,height=45mm}
\epsfig{file=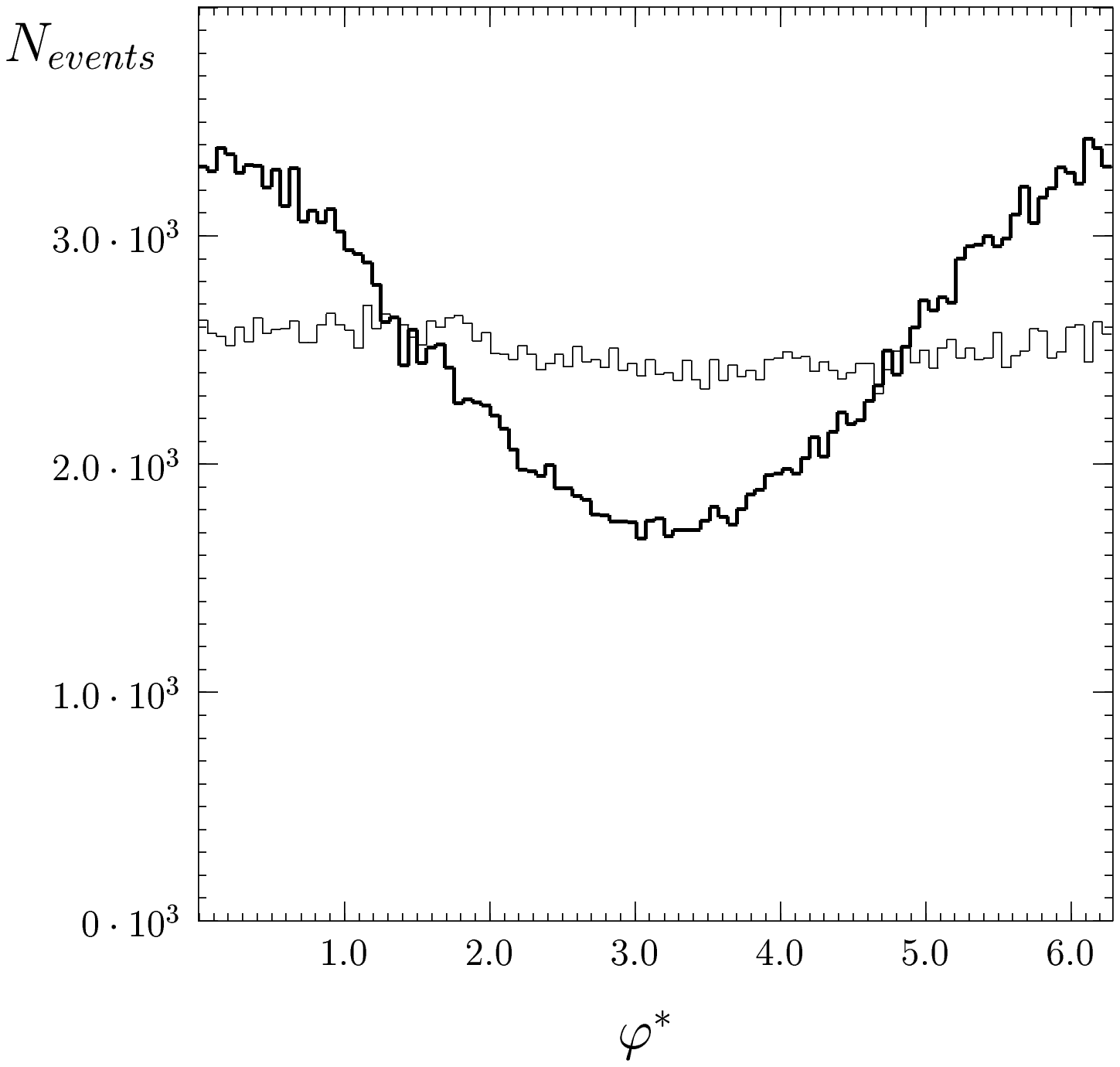,width=55mm,height=45mm}
\epsfig{file=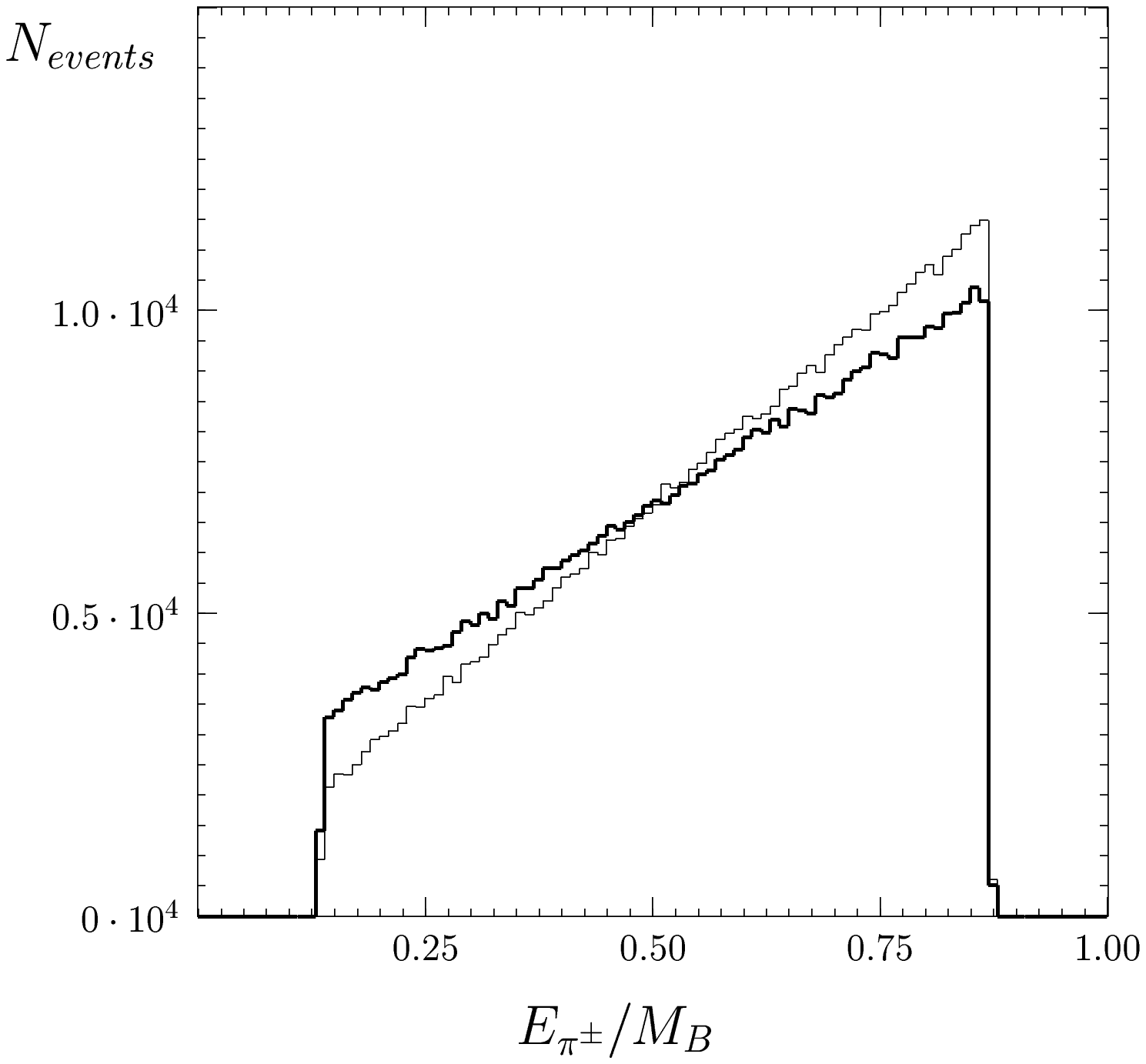,width=55mm,height=45mm}
\epsfig{file=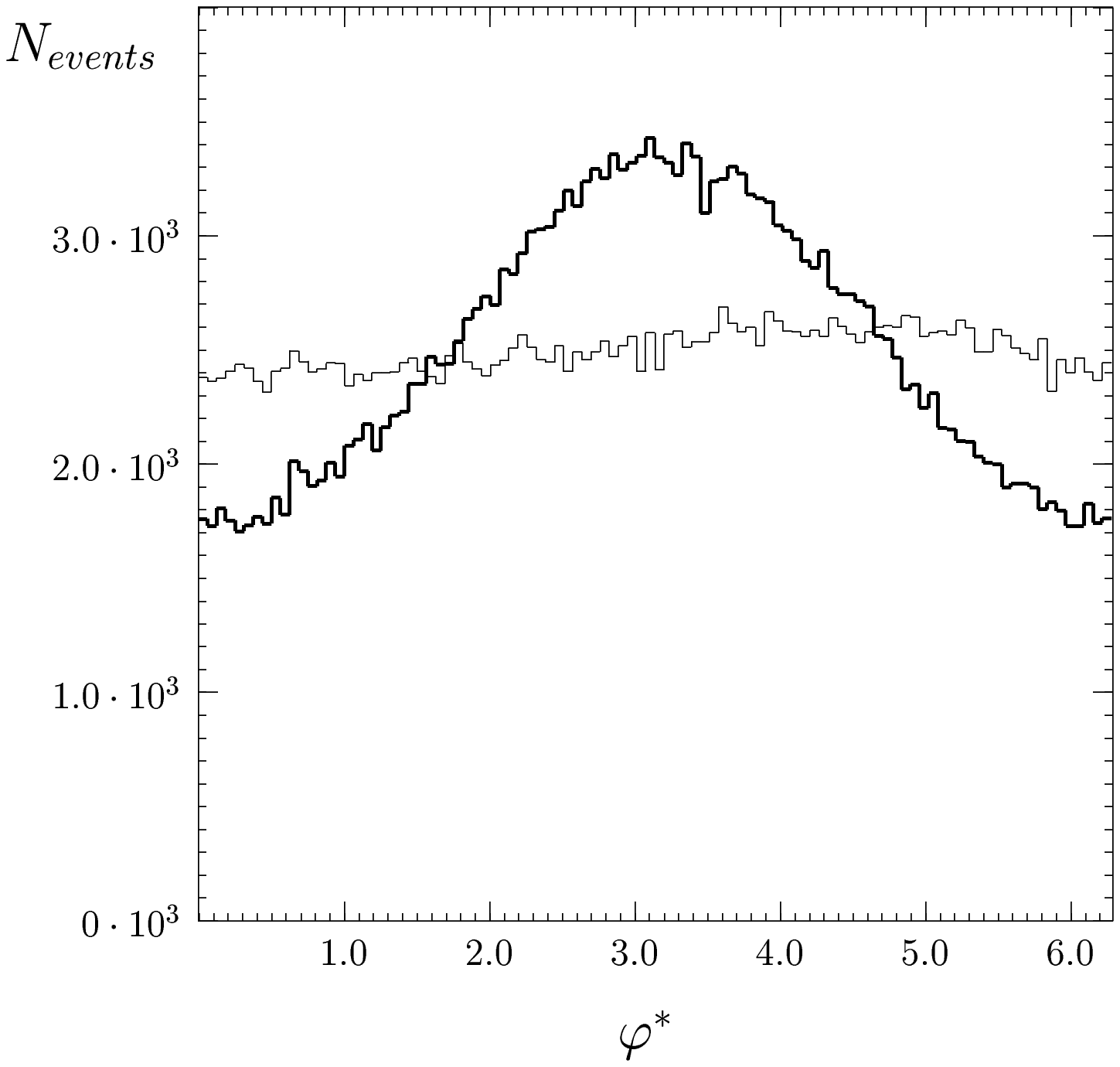,width=55mm,height=45mm}
\end{center} 
\caption  
{\it Results for the CP violating phase $\delta_{\rm CP}=0.7$ and $a=b$. 
Left panels:
Single $\pi^\pm$ energy spectra in $B^0(\bar B^0)\to\tau^+\tau^-$,
$\tau^\pm\to\pi^\pm\nu_\tau(\bar\nu_\tau)$. In the upper (lower) panel the 
thick line corresponds to the energy spectrum of $\pi^-$ (of $\pi^+$) from 
$B^0$ decays and the thin line to the energy spectrum of $\pi^+$ (of $\pi^-$) 
{}from $\bar{B}^0$. 
Right panels: acoplanarity distributions of the $\rho^+\rho^-$ decay products
in $B^0(\bar B^0)\to\tau^+\tau^-$, 
$\tau^\pm\to\rho^\pm\nu_{\tau}(\bar{\nu}_{\tau})$, 
$\rho^\pm\to\pi^\pm\pi^0$. The thick  lines correspond to the 
acoplanarity angle $\varphi^\ast$ measured in $B^0$ decays and the thin ones 
are for the angle $2\pi-\varphi^\ast$ measured in $\bar{B}^0$ decays. 
Events in the upper (lower) panel have $y_1 y_2 > 0$ ($y_1 y_2< 0$). } 
\label{fig:delta07}  
\end{figure}  
%*******************************************************************

Fig.~\ref{fig:delta07} shows the pion energy spectra and the acoplanarity 
distributions assuming $|q/p|=1$, $a=b=10^{-9}$ and the CP violating phase 
$\delta_{\rm CP}=0.7$. For all plots the same number of $5\times10^5$ 
$\tau^+\tau^-$  events from $B_d^0$ and $\bar B_d^0$ decays has been 
generated with {\tt TAUOLA}, although for the parameters chosen the ratio 
$R_\tau= 1.32$, see Fig.~\ref{fig:Rl}. In the upper left panel the energy 
spectra of $\pi^-$ from $B^0$ decays (thick line)
and of $\pi^+$ from 
$\bar B^0$ (thin line) are shown, 
while in the lower left panel shown are the spectra of $\pi^+$ from $B^0$ 
decays (thick line) and of $\pi^-$ from 
$\bar{B}^0$ (thin line).  
The harder $\pi^-$ energy spectrum from $B_d^0$ decays than $\pi^+$ from 
$\bar B_d^0$ in the upper left 
panel indicates that 
$Br(B_d^0\to \tau^+_R\tau^-_R)>Br(\bar{B}_d^0\to \tau^+_L\tau^-_L)$, 
which is a clear signal of CP violation. In the acoplanarity plots (right 
panels) thick lines correspond to the distribution of $\varphi^\ast$ measured 
in $B^0$ decays, and the thin lines to the distribution of  
$2\pi-\varphi^\ast$ measured in $\bar{B}^0$ decays; in the upper right panel  
$y_1y_2>0$, and  $y_1y_2<0$ in the lower right one. 
Different shapes of thick and thin lines seen in the right panels of
Fig.~\ref{fig:delta07} again indicate CP violation. In both energy and 
acoplanarity plots the CP violation is clearly seen and should be measurable 
even for small statistics.
Note also that if upper and lower plots are combined, ({\it i.e.} no 
sorting according to the pion charge or sign of $y_1y_2$ is made), 
all CP asymmetries are lost.

With decreasing $|\delta_{\rm CP}|$ the signal of CP
violation deteriorates (especially in the pion spectra)
and the possibility of
distinguishing  pion spectra and
acoplanarity distributions from $B^0$ and $\bar{B}^0$, 
and hence the CP violation, would require increasingly
large statistics which may not be attainable at BELLE and \babar\
without major upgrades.

\begin{figure}[!ht]
\begin{center}  
\epsfig{file=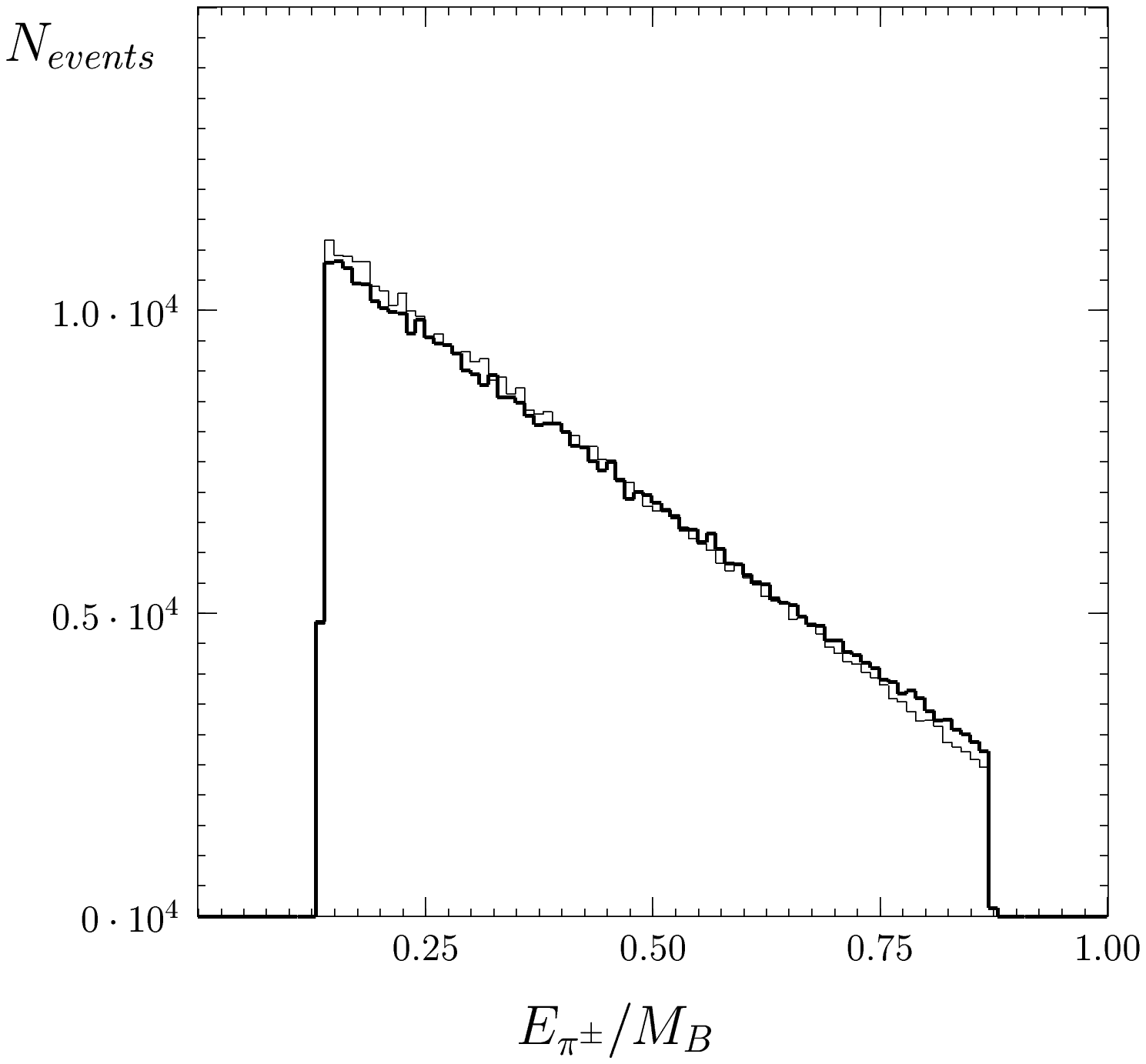,width=55mm,height=45mm}
\epsfig{file=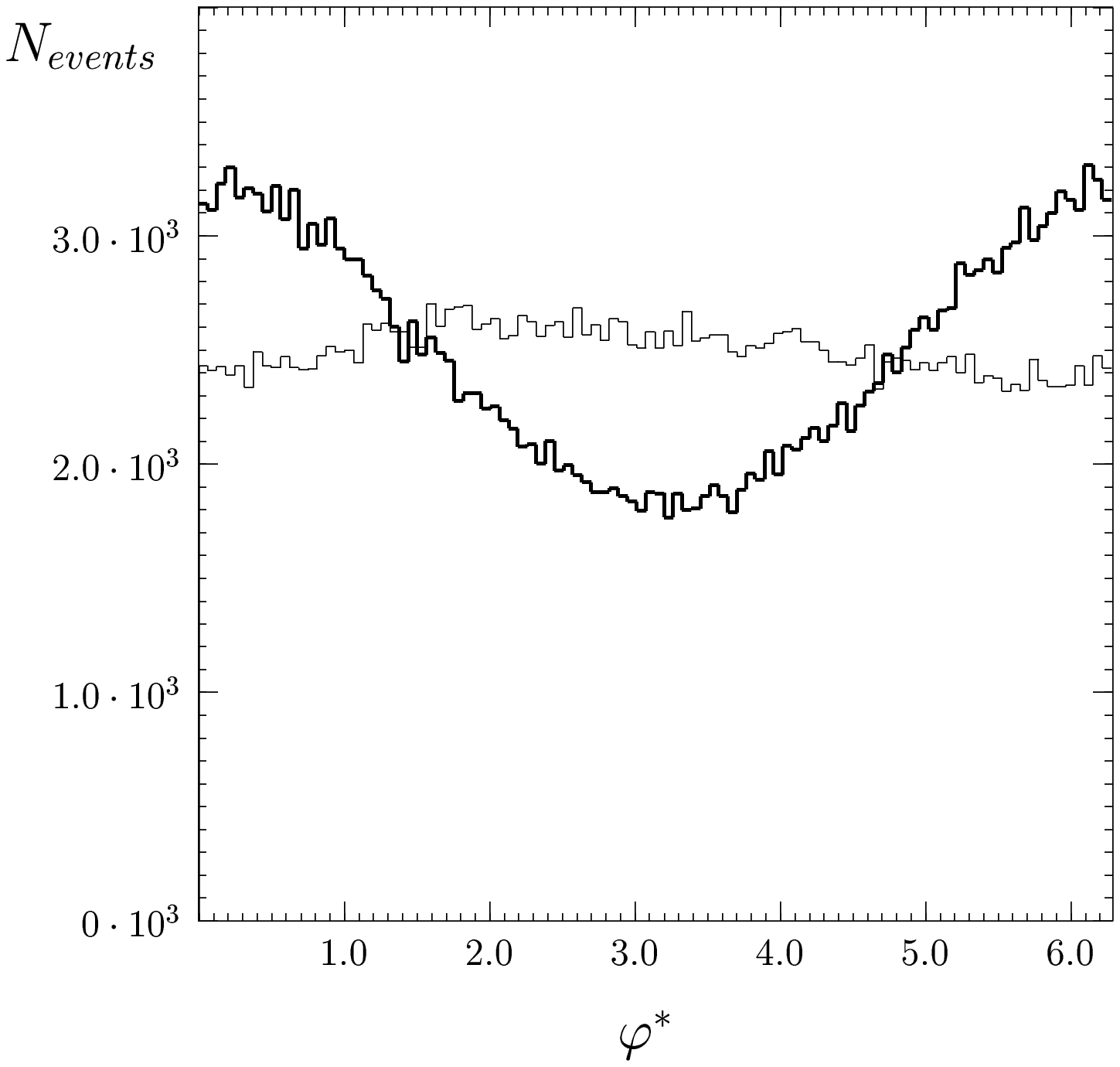,width=55mm,height=45mm}
\end{center} 
\caption 
{\it As in the upper panels of 
figure \ref{fig:delta07} but for $\delta_{\rm CP}=0.7$ 
and $a=0.8\, b$.} 
\label{fig:altb}  
\end{figure}

As we discussed, the relation $a=\pm b$ is only approximate. 
For $b=0.8~a$ with the same value 
of $\delta_{\rm CP}$ the CP violating effects
in $\pi^\pm$ energy spectra get enhanced, while the acoplanarities  
are only slightly affected.
On the other hand, for $a$ approaching $b\beta$ (for $B\to\tau^+\tau^-$ 
decays $\beta\approx0.74$) the effects of CP violation in
the $\pi^\pm$ energy spectra disappear, as expected and seen in 
the left panel of Fig.~\ref{fig:altb}. 
In contrast, the acoplanarities shown in 
the right panel of Fig.~\ref{fig:altb} clearly indicate the CP violation
even for $a\approx b\beta$ confirming our earlier discussion 
demonstrating the complementarity of the energy and acoplanarity 
distributions as a means to detect CP violation.

%*******************************************************************  
%*******************************************************************

\section{Conclusions}

In the supersymmetry scenario 
with $\tan\beta\sim40\div50$ the rates of 
$B_d^0(\bar{B}_d^0)\rightarrow\tau^+\tau^-$ decays are enhanced and  could
be detectable in the SLAC and KEK $B$-factories. Moreover,
the effective CP violating phase needs not be small. 
Therefore the CP asymmetries
can be quite large as opposed  to the $B^0(\bar{B}^0)\rightarrow\mu^+\mu^-$ 
decays in which they are kinematically suppressed. 

By using Monte-Carlo simulations we have investigated the possible 
effects of CP violation in two realistic  experimental observables 
and demonstrated that they might be detectable if the CP violating phase
is reasonably large, {\it i.e.} ${\cal O}(1)$. We have developed the
necessary formalism and numerical tools allowing to apply the {\tt TAUOLA} 
$\tau$-lepton decay library together with its {\tt universal interface} 
to simulate fully the effects of the polarization of $\tau^+$ and $\tau^-$ 
originating from such decays.
The tools can also be 
applied to determine the upper limits
on the branching fraction of the $B^0(\bar{B}^0)\rightarrow\tau^+\tau^-$
decays by the \babar\ and BELLE collaborations. \\[2mm]

\noindent {\bf Acknowledgments:}\\[1mm]  
Discussions with C.~Potter are greatly acknowledged. 
Work supported by the Polish KBN  Grants 1 P03B 099 29 (P.Ch.), 
1 P03 091 27 (Z.W.) and 1 P03B 009 27 (M.W.), and by the  
EC Contract MRTN-CT-2004-503369 "The Quest for Unification: Theory
Confronts Experiment"  (P.Ch.).
M.W. acknowledges the Maria Sk\l odowska-Curie Fellowship  
under the EU contract HPMD-CT-2001-00105 
"Multi-particle production and higher order correction".

\end{document}